\documentclass[twocolumn,letterpaper]{IEEEAerospaceCLS}  

\usepackage[]{graphicx}    
\usepackage{amssymb}
\usepackage{amsmath}
\usepackage{cancel}
\usepackage{float}
\usepackage{adjustbox}
\usepackage{multirow}
\usepackage{hyperref}
\usepackage{trackchanges}
\newcommand{\ignore}[1]{}  

\begin{document}
\title{Quantifying Operational Constraints of Low-Latency Telerobotics for Planetary Surface Operations}

\author{%
Benjamin J. Mellinkoff\\
Center for Astrophysics and Space Astronomy,\\
593 UCB, University of Colorado Boulder,\\
Boulder, CO 80309, USA\\
310-562-3928\\
benjamin.mellinkoff@colorado.edu\\
\\
Wendy Bailey\\
Engineering Management Program,\\
University of Colorado Boulder,\\
Boulder, CO 80309, USA\\
303-735-6814\\
wendy.bailey@colorado.edu
\and
Matthew M. Spydell\\
Center for Astrophysics and Space Astronomy,\\
593 UCB, University of Colorado Boulder,\\
Boulder, CO 80309, USA\\
970-291-1251\\
matthew.spydell@colorado.edu\\
\\
Jack O. Burns\\
Center for Astrophysics and Space Astronomy,\\
593 UCB, University of Colorado Boulder,\\
Boulder, CO 80309, USA\\
303-735-0963\\
jack.burns@colorado.edu
\thanks{\footnotesize 978-1-5386-2014-4/18/$\$31.00$ \copyright2018 IEEE}              
}

\maketitle

\thispagestyle{plain}
\pagestyle{plain}

\begin{abstract}
NASA's Space Launch System (SLS) and Orion crew vehicle will launch humans to cislunar space in the early 2020’s to begin the new era of space exploration. NASA plans to use the Orion crew vehicle to transport humans between Earth and cislunar space where there will be a stationed habitat known as the ``Deep Space Gateway" (DSG). The proximity to the lunar surface allows for direct communication between the DSG and surface assets, which enables low-latency telerobotic exploration. While this exploration method is promising, the operational constraints must be fully explored on Earth before being utilized on space exploration missions. This paper identifies two constraints on space exploration using low-latency surface telerobotics and attempts to quantify these constraints. One constraint associated with low-latency surface telerobotics is the bandwidth available between the orbiting command station and the ground assets. The bandwidth available will vary during operation. As a result, it is critical to quantify the operational video conditions required for effective exploration. We designed an experiment to quantify the threshold frame rate required for effective exploration. The experiment simulated geological exploration via low-latency surface telerobotics using a modified commercial-off-the-shelf (COTS) rover in a lunar analog environment. The results from this experiment indicate that humans should operate above a threshold frame rate of 5 frames per second. In a separate, but similar experiment, we introduced a 2.6 second delay in the video system. This delay recreated the latency conditions present when operating rovers on the lunar farside from an Earth-based command station. This time delay was compared to low-latency conditions for teleoperation at the DSG ($\leq$0.4 seconds). The results from this experiment show a 150\% increase in exploration time when the latency is increased to 2.6 seconds. This indicates that such a delay significantly complicates real-time exploration strategies.
\end{abstract}

\tableofcontents

\section{Introduction}
\subsection{The New Era of Space Exploration}
NASA will begin sending humans beyond the confines of LEO to expand the human presence in our solar system early in the next decade \cite{GOV}. The Space Launch System and the Orion crew vehicle will transport humans to destinations including cislunar space. These exploration missions will be used to better understand the effects of the deep space environment on humans in preparation for very long duration missions (e.g. Mars) \cite{IAA}.\par

While the Orion crew vehicle is human-rated for upwards of 21 days \cite{ECLSS}, NASA would like to establish a stationed crew habitat in cislunar space to accommodate longer duration missions. The DSG would serve as an access point to regions in deep space and the lunar surface \cite{DSG}. A potential location for the DSG is a halo orbit around the Earth-Moon L2 Lagrange Point $\sim$65,000 km above the lunar farside. This location allows for low-fuel transport between different Lagrange Points, while also providing constant line-of-sight between the Earth, lunar farside, and Sun. This line-of-sight is very appealing, as it provides direct communications with Earth while maintaining a constant vantage point over the unexplored lunar farside. In addition, the DSG stationed in orbit around the Earth-Moon L2 Lagrange Point allows for constant solar energy collection for high-power operations. While the DSG will not be manned year-round, the DSG will also serve as a communication relay between ground assets on the lunar farside and Earth command stations \cite{IAA}.\par

The DSG will promote extended exploration science missions in cislunar space and beyond \cite{DSG}, providing an opportunity to utilize low-latency surface telerobotics to remotely conduct science on the lunar farside while in orbit. Low-latency surface telerobotics is best conducted within approximately 60,000 km from the lunar surface asset to be within the human cognitive threshold for real-time perception. This threshold equates to a maximum latency of approximately 0.4 seconds \cite{cogThreshold}. This level of latency is acceptable because humans have a minimum response time of approximately 0.2 seconds \cite{tomSheridan}. This minimum response time suggests that humans cannot distinguish between real-time feedback and feedback with time delays upwards of approximately half a second. The DSG stationed at the Earth-Moon L2 Lagrange Point fulfills the requirements to achieve real-time operations between astronauts and assets on the lunar farside. The combination of the DSG and a low-latency surface telerobotics module provides a telepresence on the lunar farside \cite{humanPres}.

\subsection{Low-Latency Surface Telerobotics}
Currently, planetary telerobotics occur at a high latency; the Curiosity Rover operates with a delay from 8.6 minutes to 40 minutes \cite{HERRO}. This delay requires large amounts of time and manpower to analyze and calculate every move the rover makes. Such a valuable ground asset cannot encounter a mission-threatening problem when the delay of communication is so high. As astronauts venture beyond LEO to planetary bodies within the solar system, low-latency communications present the opportunity to establish a telepresence which was previously impossible \cite{humanPres}. Telepresence on planetary bodies provides a valuable tool for exploration. While there is always value with manpower on Earth to analyze vast amounts of data and make calculated decisions, the addition of astronauts' ability to respond in real-time when conducting missions significantly increases mission value and success \cite{HERRO}. This means keeping astronauts on board the DSG within the real-time cognitive threshold will offer increased value and success to future missions.\par

Utilization of low-latency surface telerobotics on the lunar surface will allow scientific exploration of the surface and deployment of surface infrastructure while astronauts remain in orbit \cite{LFRA}. Permanently shadowed craters on the Moon are time sensitive areas to explore due to extremely low temperatures. The use of a low-latency system would be advantageous because real-time decisions made by humans promote more science collection within shorter time periods \cite{ISS}.\par

Low-latency surface telerobotics also provides real-time human supervision with autonomous assembly, deployment, and exploration. For instance, a rover could autonomously deploy the material for a low-frequency radio array while humans supervise and rapidly correct any mistakes made by the autonomous deployment system \cite{LFRA}. It is critical to have this capability because a simple mistake made by an autonomous system when deploying or assembling can lead to mission failure.\par

The Servicing Missions for Hubble utilized astronaut EVAs to provide real-time human presence during operations. This human presence allows for improvised solutions to unexpected issues during assembly/construction. For instance, the EVA crew on STS-61 (Servicing Mission 1) had difficulty closing the latches on doors that house Hubble's gyroscopes and had to improvise to find a solution \cite{HUBBLE}. While the anomalies encountered over the course of the Hubble Servicing Missions during assembly/construction were not due to errors with an autonomous system, similar types of anomalies could easily occur with autonomous operations. Therefore, the ability for humans to improvise in real-time provides the option to remediate problems that would otherwise lead to mission failure had it been purely robotic with limited human intervention.

\begin{figure*}[h]
\centering
\includegraphics[width=\textwidth]{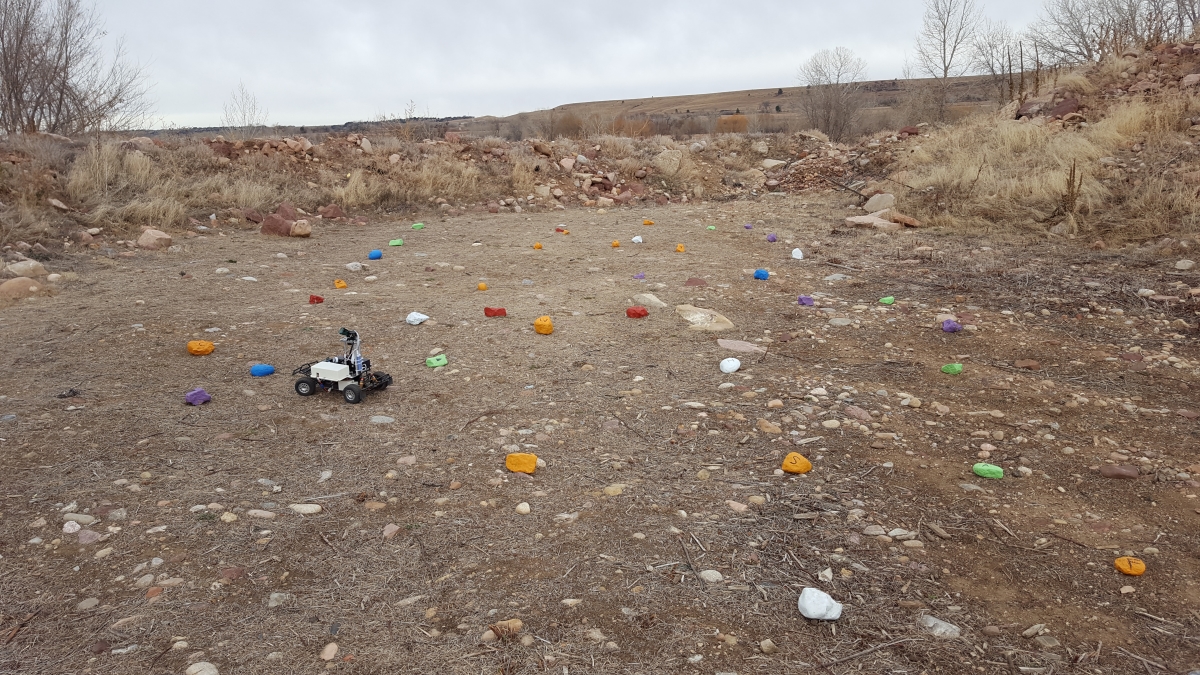}
\caption{\textbf{Operator exploring the course on the University of Colorado-Boulder's South Campus via the rover. The course was surrounded by a rim of debris and rocks similar to a crater. Some of the painted targets the operator had to identify are seen scattered on the ground.} \label{RoverCourse}}
\end{figure*}

\subsection{Limitations of Telerobotics}
The benefits that occur when utilizing low-latency telerobotics for exploration in our solar system have been well researched and would increase mission value and success \cite{HERRO}. However, the amount of research investigating the potential limitations of low-latency telerobotics for space applications has been minimal to date. An experiment conducted aboard the ISS in 2013 was the first time humans have controlled a surface asset from orbit. Astronauts teleoperated NASA's K10 rover on Earth's surface with communication latency ranging from 0.5-0.75 seconds \cite{ISS}. The results from this experiment indicate that the use of a human-robotic partnership via low-latency telerobotics is a plausible method for remotely operating on planetary surfaces. This was, in large part, the first major stride toward the R\&D necessary to make low-latency telerobotics a reality for the approaching missions aboard the Orion and DSG.\par

The limitations of telepresence for scientific exploration must be fully understood prior to implementing low-latency surface telerobotics as a strategy for exploring our solar system. Simulations of telerobotic use in space should be conducted from Earth to pinpoint the constraints of telerobotic operations. Some of the constraints might include the operational light level, required bandwidth, user interface, etc.\par

The maximum bandwidth available for communications between Earth-Moon L2 and the lunar farside is approximately 4 Megabits per second (Mbps) (assuming a 0.5 meter Ka-band antenna on the rover with 10 W of power) \cite{ISS}. However, the actual bandwidth available will sometimes drop below the 4 Mbps due to line-of-sight variability between the rover and the astronauts' communication antenna. The bandwidth is used to transmit many parts of a video feed: colorscale, resolution, and frame rate.\par

\subsection{The Bandwidth Experiment}
We ran several bandwidth and latency experiments using a COTS rover remotely controlled by a human operator in an isolated location. The human operator was tasked with identifying exploration targets in an attempt to recreate telerobotic geological exploration.\par

Our first bandwidth experiment was brief and provided an overview of the effect of frames per second (FPS) on exploration of an unfamiliar environment. The FPS, together with the video resolution and color-scale, can be used to approximate the bandwidth required for a particular video feed. A derivation of this bandwidth approximation is included in the Appendix. This overview of the effect of FPS on exploration allowed us to construct an experiment that focused our efforts of research in the correct FPS region to find meaningful results. In particular, we found that the minimum frame rate was in the range of 4-6 FPS.\par

After narrowing the FPS range we designed an experiment to quantify one factor that influences the effectiveness of telerobotics, namely  the effects of lowered bandwidth. In particular, we sought to quantify the minimum operational FPS required for a human operator to successfully explore an unfamiliar environment using low-latency surface telerobotics. To begin identifying the operational limits associated with low-latency telerobotic exploration our research group developed an experiment to investigate limitations in exploration due to reduced video frame rate. The experiment consisted of human operators remotely controlling a rover in search of ``interesting" objects (targets). The operators searched for the targets using only the video feedback provided by the rover's two cameras. The frame rate was randomly varied for each trial and the time to discovery was used as the metric of success. The modified COTS rover and the course are shown in Figure \ref{RoverCourse}.

\subsection{The Latency Experiment}
We also conducted an experiment that investigated the effect of increased latency conditions. Specifically, we examined the effect of 2.6 second latency on exploration efficiency. 2.6 seconds was chosen because it is the best case round-trip communication time from Earth to a communication satellite at Earth-Moon L2 to the lunar farside. This is the best case latency because it only accounts for time delays due to the speed of light.\par

This experiment was conducted in the same manner as the previous with a slight variation. The FPS of the video feed was fixed at 5 FPS and the introduced latency was toggled between $\leq$0.4 seconds and 2.6 seconds. Other than the fixed FPS and variable latency the experiment was conducted with the same rover and a similar course setup.

\section{Materials and Methods}
\subsection{Experimental Method}
In this experiment we had operators identify targets that were randomly distributed within a lunar-analog course and measured the time it took to find a specified target. The targets were painted rocks with different symbols on the surface. The identification of a unique target was defined as a trial in this experiment.\par

\begin{figure}[h]
\centering
\includegraphics[width=0.48\textwidth]{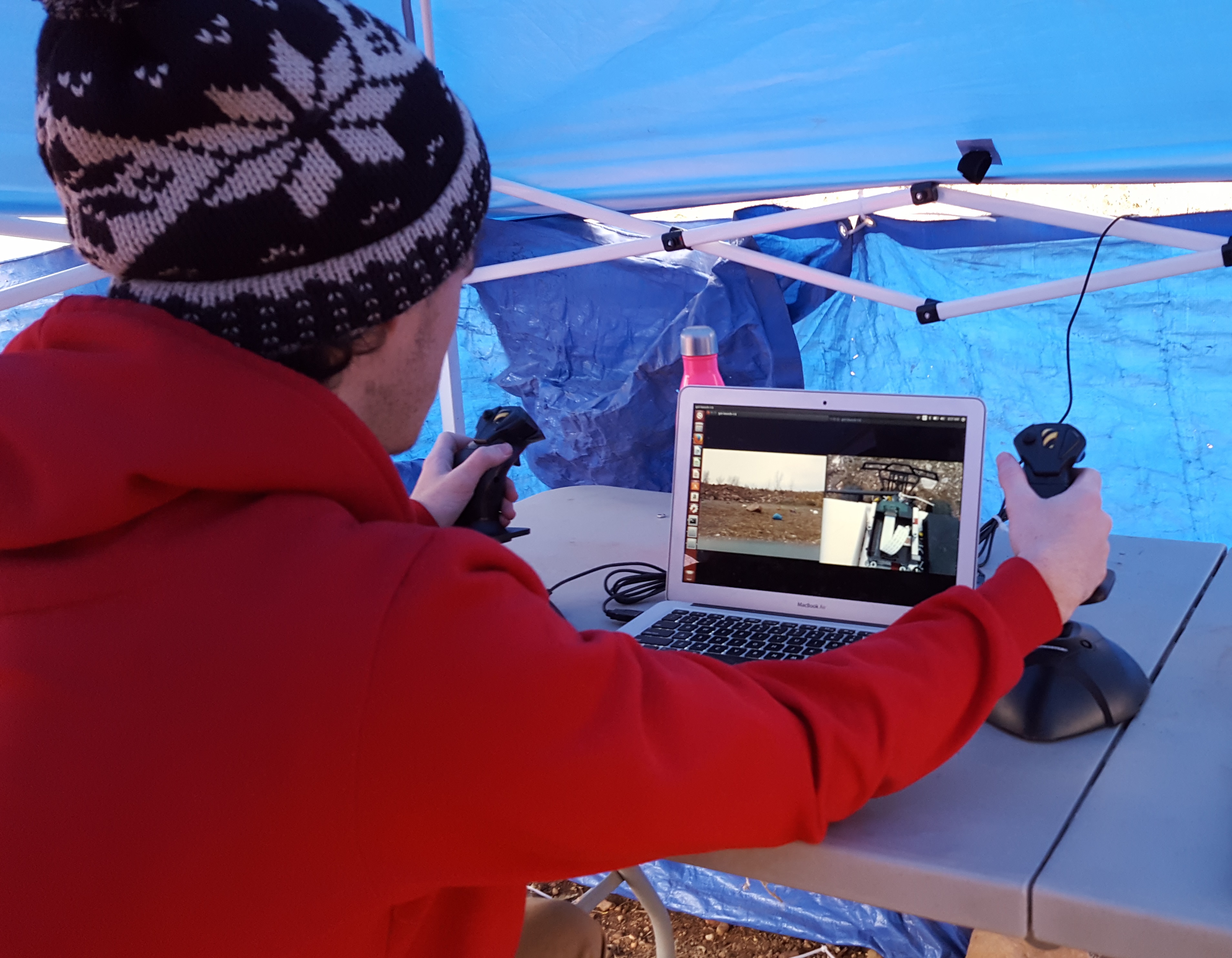}
\caption{\textbf{Operator is housed inside a three sided tent to obstruct their view of the course. The two joysticks used to manipulate the rover are on either side of the computer. The left joystick was used to control the top camera and the right joystick was used to control the rover. There are two video feeds displayed: one is from the forward facing camera, the other is from the top mounted camera.}
\label{Operator}}
\end{figure}

For the bandwidth experiment we had three different operators running the experiment and each operator ran 54 trials. The latency experiment only required two operators and 44 trials. Each trial corresponded to a search/exploration of a different target. The operators were isolated within an enclosed canopy to ensure they were operating the rover ``blind." Within this isolated environment, the operators controlled the rover through a computer interface. The two video feeds sent from the rover were displayed on the computer screen. The operator used two joysticks to control the rover's movement and the orientation of its top camera. Figure \ref{Operator} shows an operator in the closed environment controlling the rover.\par

Before the operators began the official trials for data collection they underwent training by operating the rover in search of targets in a separate location. We repeated this process with each driver until the time to discovery became constant. This was done so learning effects present for each driver would be minimized before actual trials began. We also repeated the training process briefly at the start of a new testing day for each operator. This was done to ensure the operator was performing at the same level for each day of testing.\par

\begin{figure}[h]
\centering
\includegraphics[width=0.48\textwidth]{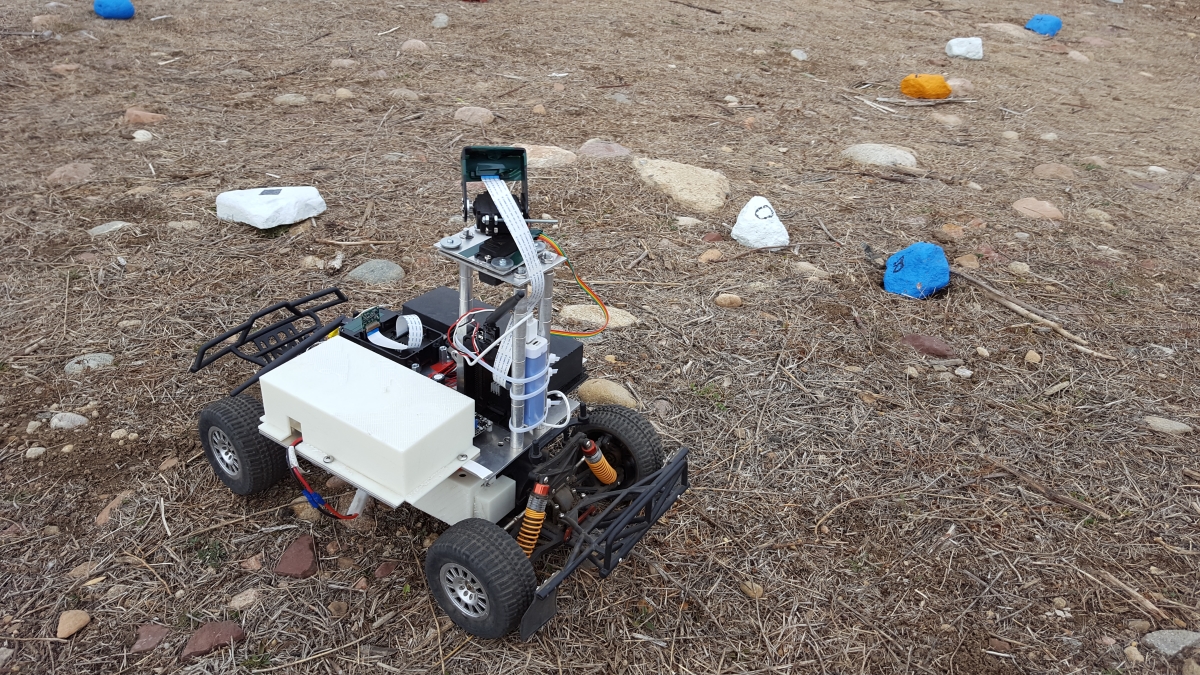}
\caption{\textbf{Close up view of the rover. The forward facing stationary camera is mounted at the front and the top camera is mounted on two servos for manipulation. The various electronics controlling the rover are housed underneath the white and black protective shields.} \label{RoverID}}
\end{figure}

The rover was located at the same starting point for each trial. Before initiating the trial we told the operator which target to discover. While there were three of each type of targets scattered in the course, the operator only had to discover one of the three targets. The timer for each trial began as soon as the operator started moving the rover. The operator explored the course in search of the target, and the timer ended after successful identification of the target object. Figure \ref{RoverID} shows the perspective from the rover when identifying a target.

\subsection{Rover}
We used a modified COTS rover for our experiment. Two new communication systems were installed; the first used a wireless router for streaming real-time video and the second used an XBee radio-frequency communication module for sending user commands. Two Raspberry Pi's and Raspberry Pi cameras were used to capture real-time video. For the bandwidth experiment we used a video service called Gstreamer to stream video from the rover to the computer \cite{gstreamer}. This service allowed us to set the frame rate and aspect ratio of the video. The latency experiment used a service called NetCat as it was easier to implement artificial latency \cite{netcat}. The wireless router used the Transmission Control Protocol (TCP) to send data packets \cite{TCP}. Using TCP ensured that data packets were received in order and without error.\par

\begin{figure}[h]
\centering
\includegraphics[width=0.35\textwidth]{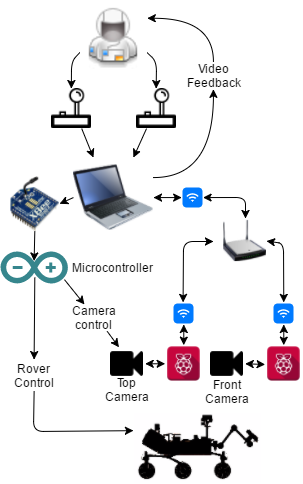}
\caption{\textbf{The user interfaces with the command computer via two joysticks. Both joystick controls are sent via XBee to a micro-controller onboard the rover. One joystick manipulates the top camera and the other manipulates the rover. Two Raspberry Pi's onboard the rover send the video feeds via WiFi and WiFi router to the command computer to be displayed for the user.} \label{diagram}}
\end{figure}

The XBee was connected to the rover with its mate connected to the operator's computer. The computer used two joysticks and a python script to create data packets that were sent over the computer's USB port to the XBee. The data packets sent to the computer's XBee were transmitted and then received by the XBee on the rover. Each data packet contained instructions to control the rover and its top camera. A micro-controller located on the rover parsed the incoming data packets and issued the instructions to the correct rover peripherals. Figure \ref{diagram} shows the flow of communications and control of our telerobotic system.

\subsection{Targets for Identification} 
The targets for identification were painted rocks with the following colors: red, green, orange, blue, purple, or white. These colors served as the broad identifier for the operator while exploring. Each object also had a symbol written on the surface in bold black lettering. The symbols were chosen from a list of letters and shapes. The following were the symbols used: `A', `B', `C', `F', `I', `J', `M', `O', `S', `X', `Square', and `Triangle'. All of the combinations of colors and symbols yielded a total of 72 unique rocks. We made 3 of each unique rock type used in this experiment making a total of 216 rocks.

\subsection{Course}
The location used for running the experiment was located on the University of Colorado Boulder's South Campus. It was chosen because of some features (e.g. crater-like bowl) that are similar to the Moon. The area of exploration used for the experiment was within the walls of a crater-like region. The crater roughly formed an L shape. One end of the course had very rocky terrain while the rest of the course had minimal to moderate rocky terrain. For the bandwidth experiment the entire crater was sectioned off into approximately 180 five foot by five foot squares using twine and stakes. Once the the entire crater was sectioned into a grid each of the 216 rocks were randomly assigned a grid location. The latency experiment had the crater sectioned into 105 seven foot by seven foot squares and a total of 132 rocks. In both cases there were more rocks than grid spaces so some of the grid locations contained more than one rock. This entire process was done so the course had a random distribution of rocks. Once all rocks were distributed the grid was removed from the course to eliminate an unnatural reference point for the operator's situational awareness.

\section{Bandwidth Experiment Results}
We sought to identify the minimum frame rate in which humans can successfully explore an unfamiliar environment using low-latency surface telerobotics. We selected the dependent variable as the ``ability to explore" and the corresponding metric as time to discovery. The treatment variable was frame rate and evaluated at the following three levels: 4, 5, and 6 FPS. A sample size of $n = 153$ was used in this bandwidth experiment.\par

The type I error used for the following statistical analyses was 5\%. 
The type I error is the probability of an incorrect rejection of a true null hypothesis. The type I error is used to compare against the p-values throughout our analysis. A p-value is the calculated probability of incorrectly rejecting the null hypothesis. It is compared against our type I error to determine whether the null hypothesis is accepted or rejected. If the p-value is less than our type I error we reject the null hypothesis.

\subsection{Testing for Normality}
\begin{figure}[h]
\centering
\includegraphics[width=0.48\textwidth]{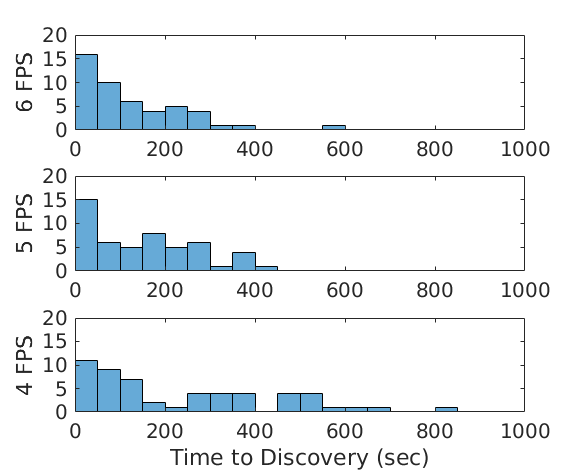}
\caption{\textbf{Visually the dispersion of time to discovery increases as frame rate decreases. Also notice the long tails for each frame rate's distribution that indicate a non-normal distribution.} \label{histEachLevel}}
\end{figure}

The first assessment of the data consisted of a test for normality of the individual treatment levels of frame rate. The moment tests for skewness and kurtosis were performed in order to validate the underlying assumptions associated with the Analysis of Variance (ANOVA) and selection of the approach for statistical evaluation of homogeneity of variances. Fig \ref{histEachLevel} shows the histograms of data by frame rate. The null hypothesis stated that the data are distributed normally, corresponding to values of Skewness ($\gamma_3$) and Kurtosis ($\gamma_4$) equal to zero. The alternative hypothesis stated that $\gamma_3$ or $\gamma_4$ do not equal zero, implying the data were not distributed normally.\par

Calculating $\gamma_3$ and $\gamma_4$ for the data within each level of frame rate using The Single-Sample Test for Evaluating Population Skewness and Kurtosis \cite{GaussianStatsTest} yields the results displayed in Table \ref{normalityTests}.

\begin{table}[h]
\centering
\textbf{\caption{\textbf{Normality Tests}\label{normalityTests}}}
\begin{adjustbox}{max width=0.48\textwidth}
\begin{tabular}{c|c|c|c|c|c}
Frame Rate & N & Skewness($\gamma_3$) & p-value & Kurtosis($\gamma_4$) & p-value \\ \hline
6 FPS & 48 & 1.545 & 1.574E-4 & 3.113 & $<$0.02 \\
5 FPS & 51 & 0.540 & 0.103 & -0.805 & $>$0.10 \\
4 FPS & 54 & 0.816 & 0.016 & -0.256 & $>$0.10    
\end{tabular}
\end{adjustbox}
\end{table}

Based on the skewness and kurtosis p-values in Table \ref{normalityTests}, we reject the null hypothesis at the 95\% confidence level and will treat the data as non-normal for the homogeneity of variance analysis. Although the treatment level of 5 FPS passed the tests for normality, we must use the approach for non-normal data based on the rejection of the null hypothesis for both 6 FPS and 4 FPS.

\subsection{Homogeneity of Variance Analysis}
An assessment for homogeneity of variance, between the treatment levels of frame rate (4, 5, and 6 FPS), was performed in order to validate the underlying assumptions associated with the ANOVA and selection of the appropriate statistical approach for any subsequent post-hoc analyses of the means. Since the data are not normally distributed we used Levene's Improved Test for Homogeneity of Variances (Brown-Forsythe) using the Absolute Deviation from the Medians (ADM) \cite{LeveneImproved}. The null hypothesis stated that the variance between the treatment levels of frame rate were equivalent. The alternative hypothesis stated that the variance of at least one of the frame rate levels was not equivalent to the others.

\begin{figure}[h]
\centering
\includegraphics[width=0.48\textwidth]{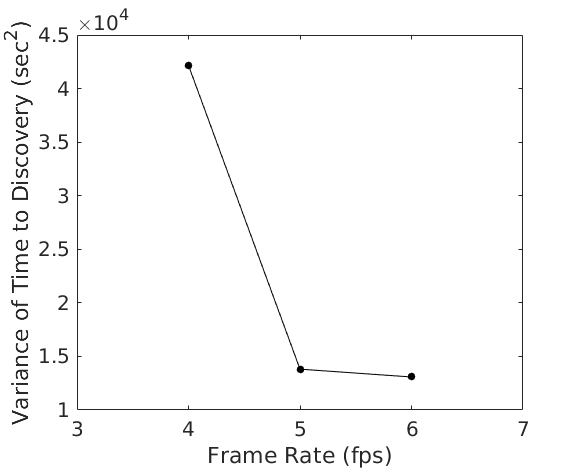}
\caption{\textbf{Plot of sample variances. The contrast between 4 FPS and 5 FPS show how unreliable exploration is at 4 FPS compared to 5 FPS. Note: lines connecting data points are included to help guide the eye.} \label{varAn_plot}}
\end{figure}

Table \ref{meansTable} lists the variance at each level of frame rate. Based on the results of the ANOVA using the ADM, the null hypothesis was rejected at the 95\% confidence level ($F(2,148)= 12.962, p = 0.000006$) and the alternative hypothesis was accepted. Figure \ref{varAn_plot} shows the plot of the variance for each frame rate.

\begin{table}[h]
\centering
\textbf{\caption{\textbf{Variance by Frame Rate}\label{meansTable}}}
\begin{adjustbox}{max width=0.48\textwidth}
\begin{tabular}{c|c|c}
Frame Rate & N & Variance \\ \hline
6 FPS & 48 & 13080.936 \\ 
5 FPS & 51 & 13812.540 \\
4 FPS & 54 & 42164.710   
\end{tabular}
\end{adjustbox}
\end{table}
Since the variance was not equivalent between frame rates we conducted a post-hoc analysis to determine which groups were different from each other. We evaluated all pairwise comparisons using the Games-Howell test on the ADM due to unequal sample sizes in each treatment level of frame rate \cite{GamesHowell}. Table \ref{stuckVarPE} shows the results of the Games-Howell post-hoc test on the sample variances between all levels of frame rate. The sample variance of 5 and 6 FPS are equivalent at the 95\% confidence level, which forms Group 1 in Table \ref{stuckVarPE}. The sample variance of 4 FPS is significantly different than 5 and 6 FPS at the 95\% confidence level, which forms Group 2 in Table \ref{stuckVarPE}.\par

These results demonstrate that the consistency of exploration and discovery at 4 FPS has more variability than at 5 and 6 FPS. The consistency of exploration is a very important factor for future telerobotic missions given the cost and importance of mission success.

\begin{table}[h]
\centering
\textbf{\caption{\textbf{Frame Rate Groups based on Variance} \label{stuckVarPE}}}
\begin{adjustbox}{max width=0.48\textwidth}
\begin{tabular}{c|c|c|c}
Frame Rate & N & Group 1 & Group 2 \\ \hline
6 FPS & 48 & 13080.94 & \\
5 FPS & 51 & 13812.54 & \\ 
4 FPS & 54 & & 42164.71                        
\end{tabular}
\end{adjustbox}
\end{table}

\subsection{Mean Time to Discovery (MTD) Analysis}

The MTD was used as the main metric for measuring the operator's ``ability to explore" as the frame rate was varied. The null hypothesis stated that the MTD ($\mu$) across all frame rates would be equal. Our alternative hypothesis stated that the MTD of at least one frame rate was not equivalent to the others.\par

\begin{figure}[h]
\centering
\includegraphics[width=0.48\textwidth]{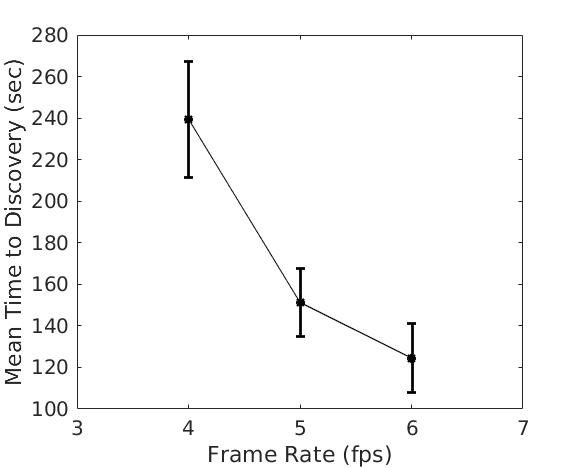}
\caption{\textbf{The MTD has a significant decrease moving from 4 FPS to 5 FPS while moving from 5 FPS to 6 FPS has minimal effect and no statistical difference. This shows a type of cutoff point moving from 5 FPS to 4 FPS. Note: vertical lines represent error bars and lines connecting data points are included to help guide the eye.} \label{margMeanPlot}}
\end{figure}

Table \ref{margMeanTable} lists the MTD calculated for each frame rate. Based on the results of the ANOVA using the MTD, the null hypothesis is rejected at the 95\% confidence level ($F(2,148)= 7.945, p = 0.000528$) and the alternative hypothesis is accepted. Figure \ref{margMeanPlot} shows the plot of estimated MTD against frame rate. Table \ref{margMeanTable} lists the values of the points shown in Figure \ref{margMeanPlot}. As shown in Figure \ref{margMeanPlot} the MTD increases with a decrease in frame rate. The large difference between 5 and 4 FPS show an accelerated deterioration of MTD of exploration in an unfamiliar environment.

\begin{table}[h]
\centering
\textbf{\caption{\textbf{Mean Time to Discovery} \label{margMeanTable}}}
\begin{adjustbox}{max width=0.48\textwidth}
\begin{tabular}{c|c|c|c|c|c}
Frame & & & & & \\
Rate & N & MTD & Std. Deviation & Low & High \\ \hline
6 FPS & 48 & 124.5208 & 114.37192 & 2.00 & 554.00 \\
5 FPS & 51 & 151.3137 & 117.52676 & 7.00 & 410.00 \\
4 FPS & 54 & 239.5370 & 205.34046 & 17.00 & 819.00    
\end{tabular}
\end{adjustbox}
\end{table}

Since the MTD were not equivalent between frame rate levels, we conducted a post-hoc analysis to determine which groups were different from each other. We evaluated all pairwise comparisons using the Games-Howell test on the ADM due to unequal variances and unequal sample sizes in each treatment level of frame rate \cite{GamesHowell}. Table \ref{meanPointTable} shows the results of the Games-Howell post-hoc test on the MTD between all levels of frame rate. The MTD of 5 and 6 FPS are equivalent at the 95\% confidence level, which forms Group 1 in Table \ref{meanPointTable}. The MTD of 4 FPS is significantly higher than 5 and 6 FPS at the 95\% confidence level, which forms Group 2 in Table \ref{meanPointTable}. The results reinforce what we see in Figure \ref{margMeanPlot} and prove there is a significant increase in the MTD from 5 to 4 FPS. Through our analysis, we ruled out the possibility of additional independent variables. We found distances to other targets and the time of day were not statistically significant in determining the MTD. This rapid deterioration in the MTD signifies that a minimum of 5 FPS is necessary for effective exploration using our operation parameters.

\begin{table}[h]
\centering
\textbf{\caption{\textbf{Frame Rate Groups based on MTD} \label{meanPointTable}}}
\begin{adjustbox}{max width=0.48\textwidth}
\begin{tabular}{c|c|c|c}
Frame Rate & N & Group 1 & Group 2 \\ \hline 
6 FPS & 48 & 124.5208 & \\
5 FPS & 51 & 151.3137 & \\ 
4 FPS & 54 & & 239.5370 
\end{tabular}
\end{adjustbox}
\end{table}

\section{Latency Experiment Results}
We sought to identify the operational difference between two latency conditions to show the difference between Earth based operations and DSG operations at Earth-Moon L2. Again, we selected the dependent variable as the time to discovery. The treatment variable was latency and evaluated at the following two levels: 0.4 and 2.6 seconds. These were the only two latency values used in this experiment so that a concise comparison could be made between teleoperating at the DSG and from Earth.\par

The type I error used for the following statistical analyses was 5\%.

\subsection{Testing for Normality}
As with the bandwidth experiment, the first assessment of the data consisted of a test for normality of the individual treatment levels of frame rate. The Anderson-Darling ($A^2$) test \cite{AndersonDarling} was used to test for normality in order to validate the underlying assumptions associated with the ANOVA and selection of the approach for statistical evaluation of homogeneity of variances. Figure \ref{latencyHist} shows the histograms of data by latency condition. The null hypothesis stated that the data follow a specified distribution and are distributed normally, corresponding to a value of $A^2$ ($A^2$ is the Anderson-Darling statistic that is calculated from the z-values of a data set) equal to zero. The alternative hypothesis stated that $A^2$ does not equal zero, implying the data were not distributed normally.\par

\begin{table}[h]
\centering
\textbf{\caption{\textbf{Normality Tests}\label{normalityTests_latency}}}
\begin{adjustbox}{max width=0.48\textwidth}
\begin{tabular}{c|c|c|c|c|c}
Latency &&& $A^2$ && \\ 
Condition & Operator & N & (Anderson-Darling) & p-value \\ \hline
0.4 seconds & 1 & 22 & 0.929 & 0.018 \\
0.4 seconds & 2 & 22 & 0.833 & 0.031 \\
2.6 seconds & 1 & 22 & 1.004 & 0.011 \\
2.6 seconds & 2 & 22 & 1.243 & 0.003
\end{tabular}
\end{adjustbox}
\end{table}

Based on the Anderson-Darling $A^2$ p-values in Table \ref{normalityTests_latency}, we reject the null hypothesis at the 95\% confidence level and will treat the data as non-normal for the homogeneity of variance analysis.

\subsection{MTD Analysis}

Again with this experiment, the MTD was used as the main metric for measuring the operator's ``ability to explore" as latency was varied. The null hypothesis stated that the MTD ($\mu$) across all latency conditions would be equal. Our alternative hypothesis stated that the MTD of at one latency condition was not equivalent to the other.\par

\begin{figure}[h]
\centering
\includegraphics[width=0.48\textwidth]{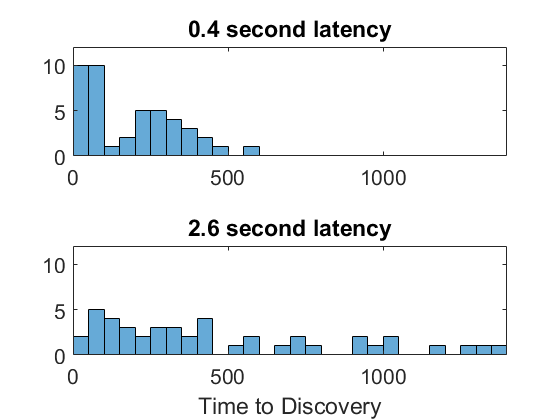}
\caption{\textbf{Visually the dispersion of time to discovery increases abruptly with the 2.6 second latency. Many trials at 2.6 seconds of latency take twice the time when compared to the maximum time at 0.4 seconds of latency.} \label{latencyHist}}
\end{figure}

The null hypothesis is rejected at the 95\% confidence level ($F(1,85) = 19.909, p = 0.000025$) and the alternative hypothesis is accepted. Table \ref{meanPointTable_latency} lists the MTD values of both latency conditions. Accepting the alternative hypothesis means the MTD of each latency condition is statistically different and table \ref{meanPointTable_latency} reflects this result.\par

\begin{table}[h]
\centering
\textbf{\caption{\textbf{Latency Groups based on MTD} \label{meanPointTable_latency}}}
\begin{adjustbox}{max width=0.48\textwidth}
\begin{tabular}{c|c|c|c}
Latency Condition & N & Group 1 & Group 2 \\ \hline 
0.4 seconds & 44 & 188 & \\
2.6 seconds & 44 & & 470 
\end{tabular}
\end{adjustbox}
\end{table}

There is a drastic change in MTD moving from 0.4 seconds of latency to 2.6 seconds ultimately resulting in a 150\% increase in MTD. Figure \ref{latencyHist} easily shows this major difference between the latency conditions visually. Again, we ruled out the possibility of additional independent variables. We found distances to other targets, starting location, and time of day were not statistically significant in determining the MTD.

\section{Related Work \& Future Work}
We reviewed previous work on the effects of video quality to validate the results from our experiment. In particular, we examined studies that investigated the effects of video quality deterioration to determine a relationship between performance and video quality for video games. While the results from each of the studies differ slightly, ultimately the studies all concluded that there is a threshold value for video quality to maintain operability. \par

The first study \cite{videogame} explored the relationship between video frame rate and resolution on a user's ability to effectively shoot an opponent in a first-person-shooter video game. They came to the conclusion that user performance is up to 7 times worse at frame rates as low as 4 FPS compared to the performance when operating at 30 FPS. In particular, they found that the video game became essentially inoperable around 4 FPS. It is important to note that the threshold frame rate is likely dependent on the speed of operations in the video game.\par

The next study \cite{Ranadive} inspected the effects of frame rate, resolution, and colorscale on an operators ability to perform a task. The tasks were specific to undersea teleoperation and included bolting/unbolting, lifting, opening/closing valves, connecting hoses, etc. The study determined that resolution and colorscale can be low while maintaining operability, while a frame rate below 5 FPS produces a considerable degradation in performance and increase in variability.\par

Studies performed within the field of telemedicine are similar to our latency experiment. An experiment conducted by the Centre of Minimal Access Surgery \cite{telemedicine} looked at the impact of varying latency on teleoperated surgical tasks. Their conclusion was that latency up to 0.5 seconds will still allow surgeons to perform tasks; however, the task completion time and number of errors goes up as higher latency is introduced. If telerobotics are to be used on planetary bodies for assembly and deployment, which require similar accurate movements as telemedicine, then human supervision should not incur latency conditions much higher than 0.5 seconds to limit errors and enforce timely error detection and correction.

The results from these studies reinforce our data and conclusion. The first and second studies indicate there is a relationship between minimum operational frame rate and the speed of the task. Future work should explore this relationship as it is relevant for future low-latency surface telerobotic missions, especially as telerobotic operation speeds increase. \par

Lastly, future work should be invested into video compression applications for telerobotics. Applying powerful video compression to telerobotics in intelligent ways will significantly lower the necessary bandwidth for telerobotic operations; it may be the key to making telerobotics an extremely powerful and robust tool for future exploration. As we have described earlier, orbits around other planets will not always produce perfect bandwidth conditions. A high tolerance of bandwidth variability is crucial for effective and continuous teleoperation missions. Video compression is a technique that can significantly increase the tolerance of bandwidth variability by drastically decreasing the worst case bandwidth required for telerobotic missions.\par

\section{Conclusion}
Our objective in the bandwidth experiment was to investigate the effects of lowered bandwidth, in particular, to find the minimum operational FPS required for a human operator to successfully explore an unfamiliar environment using low-latency surface telerobotics. Our results show a threshold for exploration discovery occurs at 5 FPS. Moving to a lower FPS causes a large jump in MTD. There are many variables that determine the exact placement and shape of the MTD curve. These variables include: FPS, resolution, colorscale, task performed, force-feedback, etc. Our data fit the trend that many other frame rate experiments produced and shows that exploring unfamiliar environments given our resolution, colorscale, and operation speed requires a minimum of 5 FPS. Therefore, as the available bandwidth between the rover and the command station drops due to variable line-of-sight, it is imperative not to operate below 5 FPS.\par

Our objective in the latency experiment was to show the difference in telerobotic operation given even a small latency. Specifically, we compared low-latency telerobotics to telerobotic operation conducted on the lunar farside from Earth. Our results show that even small amounts of latency can affect real-time operation quite drastically, thus necessitating a different operational strategy such as asynchronous control. The ability for missions to sustain low-latency operations will create a telepresence on the surface that drastically increases mission's value and success. The data support the use of humans in some capacity for future telerobotic operation; the value added to missions is worth the astronaut's time.
\appendix{}
\subsection{Calculating Bandwidth from Frame Rate}

Frame rate is an abstraction of data rate or bandwidth. Finding the minimum frame rate will ultimately produce a minimum video/visual bandwidth necessary for effective exploration. The following is an equation to calculate the video bandwidth based on FPS, resolution, and colorscale:
\begin{equation*}
    \frac{\cancel{frames}}{second} \cdot \frac{\cancel{pixels}}{\cancel{frame}} \cdot \frac{bits}{\cancel{pixel}} = \frac{bits}{second}
\end{equation*}
The first term is how many frames are sent every second. The second term is the number of pixels in each frame; this affects the resolution of the image. The last term is the colorscale which corresponds to the number of color/shade combinations possible for each pixel. After some of the terms cancel we are left with the absolute worst case bits per second for a given set of parameters. This is the worst case because no compression strategies are taken into account.\par
Using the parameters from our experiment yields:
\begin{equation*}
    \frac{5\ \cancel{frames}}{second} \cdot
    \frac{(640\times480)\ \cancel{ pixels}}{\cancel{frame}} \cdot
    \frac{24\ bits}{\cancel{pixel}} = \frac{36.864\cdot 10^6\ bits}{second}\\
\end{equation*}
\begin{equation*}
    = 36.864\ Mbps
\end{equation*}
So, at worst our video bandwidth would be $36.864$ Mbps for one video stream. This was not our effective bandwidth though. Our video stream used a popular video compression codec called H.264. This codec can compress video by approximately 70\% to 93\%, depending on the motion present in the video \cite{h264}.\par

As mentioned previously, an estimate on the maximum bandwidth available for communications between Earth-Moon L2 and the lunar far-side is approximately 4 Mbps \cite{ISS}. The data rate calculated above does not fall under the 4 Mbps limit; however, by taking into account various strategies for reducing bandwidth our 36 Mbps can be drastically reduced and brought within the threshold for lunar far-side communications.\par

The first strategy is video compression. As mentioned above, H.264, has compression ranging from 70\% to 93\%. The second strategy decreases the colorscale by using a technique that compresses a 24 bit colorscale to an 8 bit colorscale without sacrificing quality \cite{trueColor}. Using these compression factors our video bandwidth would change from 36.864 Mbps to 1.72 Mbps. \par

Besides compression techniques, the actual maximum bandwidth available between Earth-Moon L2 and the lunar farside could be increased by either increasing the size of the antenna stationed at Earth-Moon L2, increasing the power of the signal sent from the rover, or implementing an optical communication system \cite{OPTCOMM}.




\acknowledgments
Funding: This work was funded, in part by the Lockheed Martin Space Systems Company. This work was directly supported by the NASA Solar System Exploration Research Virtual Institute cooperative agreement number 80ARC017M0006. We especially thank Chris Norman, Raul Monsalve, and Terry Fong for providing their input on this project.

\bibliographystyle{IEEEtran}
\bibliography{main}

\thebiography
\begin{biographywithpic}
{Benjamin Mellinkoff}{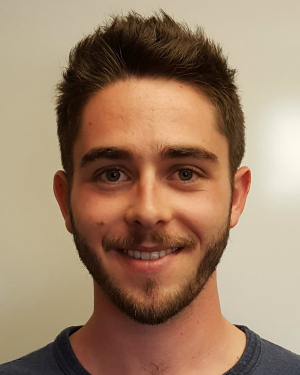}
is an undergraduate studying Aerospace Engineering Sciences at the University of Colorado-Boulder. He is currently pursuing a BS/MS degree with an emphasis in Aerospace Systems: Controls. He is the manager of the Center for Astrophysics and Space Astronomy's undergraduate telerobotics group. He is also a member of NASA/SSERVI's Network for Exploration and Space Science (NESS) team.

\end{biographywithpic} 

\begin{biographywithpic}
{Matthew Spydell}{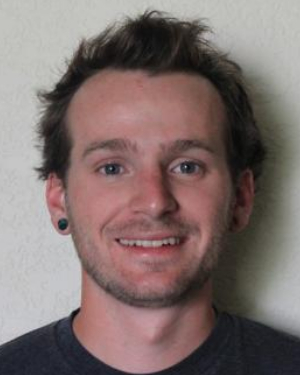}
is an undergraduate studying electrical engineering and computer engineering at the University of Colorado-Boulder. He is the electrical and computer engineer for the Center for Astrophysics and Space Astronomy's undergraduate telerobotics group. He is also a member of NASA/SSERVI's Network for Exploration and Space Science (NESS) team.

\end{biographywithpic}

\begin{biographywithpic}
{Wendy Bailey}{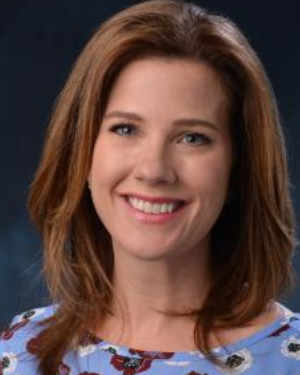}
is an Instructor in the Lockheed-Martin Engineering Management Program, teaching in the area of Quality Science. She earned her undergraduate degree in Mechanical Engineering from Purdue University, and a Masters of Engineering from the Lockheed-Martin Engineering Management Program at the University of Colorado Boulder, with an emphasis in six sigma, quality systems and applied statistics. Prior to her graduate degree, she was trained in statistical methods by Luftig \& Warren International (LWI). Wendy also worked for 14 years at Anheuser-Busch, where she became skilled in the application of statistics in an industrial environment.

\end{biographywithpic}

\begin{biographywithpic}
{Jack O. Burns}{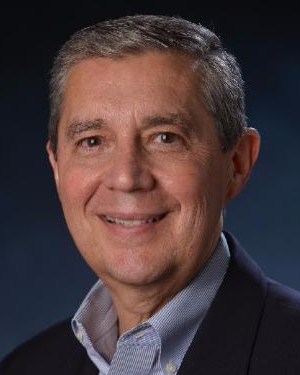}
is a Professor in the Department of Astrophysical and Planetary Sciences and Vice President Emeritus for the University of Colorado. He is also Director of the NASA-funded SSERVI Network for Exploration and Space Science (NESS). Burns is an elected Fellow of the American Physical Society and the American Association for the Advancement of Science. He received NASA's Exceptional Public Service Medal in 2010 and NASA's Group Achievement Award for Surface Telerobotics in 2014. Burns was a member of the Presidential Transition Team for NASA in 2016/17. Burns recently served as senior Vice President of the American Astronomical Society.

\end{biographywithpic}

\end{document}